\documentstyle[preprint,aps,eqsecnum]{revtex}

\def\beq#1{\begin{equation} \label{#1}}
\def\eeq{\end{equation}}

\def\ket#1{\left\vert #1\right\rangle}

\newskip\humongous \humongous=0pt plus 1000pt minus 1000pt
\def\caja{\mathsurround=0pt}

\newif\ifdtup
\def\panorama{\global\dtuptrue \openup1\jot \caja
        \everycr{\noalign{\ifdtup \global\dtupfalse
        \vskip-\lineskiplimit \vskip\normallineskiplimit
        \else \penalty\interdisplaylinepenalty \fi}}}
\def\eqalignno#1{\panorama \tabskip=\humongous
        \halign to\displaywidth{\hfil$\displaystyle{##}$
        \tabskip=0pt&$\displaystyle{{}##}$\hfil
        \tabskip=\humongous&\llap{$##$}\tabskip=0pt
        \crcr#1\crcr}}
\begin{document}
{
\tighten
\title{Pentaquark Update After Ten Years}

\author{Harry J. Lipkin\,$^{a,b}$\thanks{Supported
in part by grant No. I-0304-120-.07/93 from The German-Israeli Foundation
for Scientific Research and Development}}

\address{ \vbox{\vskip 0.truecm}
  $^a\;$Department of Particle Physics \\
  Weizmann Institute of Science, Rehovot 76100, Israel \\
\vbox{\vskip 0.truecm}
$^b\;$School of Physics and Astronomy \\
Raymond and Beverly Sackler Faculty of Exact Sciences \\
Tel Aviv University, Tel Aviv, Israel}

\maketitle

\begin{abstract}%

Studying scattering of heavy flavor hadrons and looking for bound states is
shown to give experimental information otherwise unobtainable about effective
two-body interactions between constituent $(qq)_6$ and $(\bar qq)_8$ pairs
respectively in color sextet and color octet states. All the successes of the
constituent quark model in $(uds)$ hadron spectroscopy are shown to depend only
on effective two-body interactions in color $3^*$ and singlet states.

New directions for bound pentaquark searches are discussed following the
availability of vertex detectors which can pinpoint events where a proton is
emitted from a secondary vertex. Any such event indicates a particle decaying
weakly by proton emission and the discovery of a new particle if its mass is
higher than that of known charmed baryons. There is no combinatorial
background and striking decay signatures like $p \phi \pi^-$ are no longer
needed.

The beauty pentaquark  ($\bar b suud$) and the doubly-strange pentaquark
($\bar c ssud$) may be relevant to future searches. A simple calculation shows
that the effects of flavor-SU(3) breaking on their binding relative to the
relevant thresholds are similar to that for the singly-strange pentaquark
($\bar c suud$) relative to the $D_s p$ threshold.

\end{abstract}%

} % end tighten

\section{Introduction }

It is now ten years since the proposal of the existence of the
pentaquark\cite{Moriond} and
suggestions for its search via the $p \phi \pi^-$ decay
mode\cite{Houches,LipKEK,PANIC,Pentaqua,Gignoux,Pstanford,Sivers}.
Despite subsequent experimental progress and analysis of models for pentaquark
structure, decay and signatures\cite{PSTGOAR}
there is today still no convincing experimental evidence for the existence
of the pentaquark nor of any other exotic hadron which cannot be described in a
constituent quark model as a $3q$ or $\bar qq$ state\cite{Sharon,E791Col}.
Not a single theoretical prediction for the multiquark sector has been
confirmed by experiment \cite{Sivers,Lipk86}.

    Why should anyone look for an anticharmed strange baryon?
Why should it be bound? Who cares whether it is bound?
In trying to understand how QCD makes hadrons out of quarks and gluons,
we note that long-standing regularities and paradoxes in simple experimental
hadron physics still remain to be explained by QCD.
Today the simple Levin-Frankfurt additive quark model prediction
\cite{LevFran,LS} still fits experimental data up to 310 Gev/c\cite{PAQMREV}
with a discrepancy always less than 7\% for the prediction
$$ \delta_{AQM} \equiv
(2/3)\cdot \sigma_{tot}(pp) - \sigma_{tot}(\pi^-p) \approx 0
\eqno(WW1) $$
There seems to be underlying dynamics
describing the meson-baryon difference primarily in terms of the number of
constituent quarks.
The pion is still 2/3 of a proton
even though some theorists say that the pion is a
Goldstone boson and the proton is a skyrmion. Further evidence that mesons  and
baryons are made of the same quarks is given by the remarkable successes of the
constituent quark model, in which static properties and low lying excitations
of both mesons and baryons are described as simple composites of asymptotically
free quasiparticles having an effective mass with exactly the same value for
predicting hadron masses, magnetic moments and hyperfine splittings.

The mass difference between strange and nonstrange quarks $m_s-m_u$ is found to
have the same value $\pm 3\%$
when calculated in two independent ways from baryon masses
and in two independent ways from meson
masses\cite{SakhZel,ICHJLmass,HJLMASS},
$$
\langle m_s-m_u \rangle_{Bar}= M_\Lambda-M_N=177\,{\rm MeV}=
{{M_N+M_\Delta}\over 6}\cdot
\left({{M_{\Delta}-M_N}\over
{M_{\Sigma^{\scriptstyle *}}-M_\Sigma}} - 1 \right)
=190\,{\rm MeV}.\eqno(WW2a)
$$
$$
\langle m_s-m_u \rangle_{mes} =
{{3(M_{K^{\scriptstyle *}}-M_\rho )
+M_K-M_\pi}\over 4} =180\,{\rm MeV} = $$ $$
= {{3 M_\rho + M_\pi}\over 8}
\cdot
\left({{M_\rho - M_\pi}\over{M_{K^*}-M_K}} - 1 \right)
= 178\,{\rm MeV}.\eqno(WW2b)
$$
The same approach applied to the mass difference between $b$ and $c$ quarks
$m_b-m_c$ gives
$$
\langle m_b-m_c \rangle_{Bar}= M(\Lambda_b)-M(\Lambda_c) =3356 \,{\rm MeV},
\eqno(WW2c)
$$
$$
\langle m_b-m_c \rangle_{mes} =
{{3(M_{B^{\scriptstyle *}}-M_{D^{\scriptstyle *}})
+M_B-M_D}\over 4} =3338 \,{\rm MeV}.\eqno(WW2d)
$$
The ratio ${{m_s}\over{m_u}}$ has the same value $\pm 2.5\%$
for mesons and baryons.
$$ \left({{m_s}\over{m_u}}\right)_{Bar} =
{{M_\Delta - M_N}\over{M_{\Sigma^*} - M_\Sigma}} = 1.53 =
\left({{m_s}\over{m_u}}\right)_{Mes} =
{{M_\rho - M_\pi}\over{M_{K^*}-M_K}}= 1.61
\eqno(WW3) $$
We also note three predictions of hadron magnetic
moments with no free parameters\cite{DGG,Protvino}
$$
\mu_\Lambda=
-0.61
\,{\rm n.m.}=
-{\mu_p\over 3}\cdot {{m_u}\over{m_s}} =
-{\mu_p\over 3} {{M_{\Sigma^*} - M_\Sigma} \over{M_\Delta - M_N}}
=-0.61 \,{\rm n.m.}
\eqno(WW4) $$
$$
-1.46 =
{\mu_p \over \mu_n} =
-{3 \over 2}
\eqno(WW5a)
$$
$$
\mu_p+\mu_n= 0.88 \,{\rm n.m.}
={M_{\scriptstyle p}\over 3m_u}
={2M_{\scriptstyle p}\over M_N+M_\Delta}=0.865 \,{\rm n.m.}
\eqno(WW5b)
$$
As long as
QCD calculations have not yet succeeded to explain these striking experimental
facts, it is of interest to search for new experimental input.

Additional input is obtainable from investigations of two striking
features of the hadron spectrum: (1)
the absence of strongly bound multiquark exotic
states like a dipion with a mass less than two pion masses or a dibaryon
bound by 100 MeV; (2) the structure of nuclei as composed of three-quark
clusters called nucleons rather than a quark gas, quark bags, a quark
shell model \cite{Igal,Arima} or a quark-gluon plasma.
The constituent quark model gives a very simple answer\cite{WhyNuc}.
But the validity of this simple picture still remains to be confirmed by
experiment.

All the successes of the constituent quark model with a two-body color-exchange
interaction\cite{TriEx,DGG,ICHJLmass,HJLMASS} and all of hadron spectroscopy
without exotics including scattering still give information only about the
$(\bar qq)_1$ interaction in the color singlet state and the $(qq)_{3*}$
interaction in the color antitriplet state and no information about
short-range $(qq)_6$ or $(\bar qq)_8$ interactions in color sextet or color
octet states.
The two-body color-exchange color-electric interaction commonly used
saturates\cite{TriEx} and gives no forces between two color singlet hadrons.
The two-body color-exchange color-magnetic interaction is always repulsive
between two quarks of the same flavor. Thus baryon-nucleon scattering in the
(u,d,s) sector is dominated by a short-range repulsion (the well-known
repulsive core in the nucleon-nucleon interaction). Meson-hadron scattering in
the (uds) sector must have either a quark or an antiquark in the meson with the
same flavor as one of the quarks in the other hadron. Two quarks of the same
flavor have a repulsive color-magnetic interaction keeping the two hadrons
apart. A $\bar qq$ pair of the same flavor can annihilate and produce a hadron
resonance. Thus hadron-hadron scattering in the (u,d,s) sector is dominated
either by $qq$ repulsion or by resonances produced by $\bar qq$ annihilation.

Only when there are more than three flavors is it possible to have realistic
scattering experiments (unrealistic cases like $K^- \Delta^-$ and $\phi N$
are excluded) in which there is no common flavor between beam and target
and the $(qq)_6$ or $(\bar qq)_8$ interactions can be observed.

The possible existence of exotic hadrons
remains a principal question in hadron spectroscopy
and the understanding of how the binding of quarks and gluons into hadrons is
described by QCD\cite{Lipk86}.
The first exotic hadron search was for the $H$ dibaryon\cite{Jaffe}.
Jaffe's original calculation and subsequent work \cite{Rosner}
indicate a gain in hyperfine interaction energy by recoupling color and spins
in the six quark system over the two-$\Lambda$ system. But a lattice gauge
calculation\cite{Thacker} indicated the $H$ to be unbound and well above the
$\Lambda \Lambda$ threshold. The lattice calculation showed
a repulsive $\Lambda$-$\Lambda$ interaction generated by quark exchange
\cite{Pstanford,Thacker2} which is not included in simple model calculations
and could well prevent the six quarks from coming close enough together to
feel the additional binding of the short range hyperfine interaction.

Such a repulsive exchange force can not be present in pentaquarks, which were
shown to have a hyperfine binding roughly equal to that of the $H$, but
with no possible quark exchange force in the lowest decay channel
$D_sN$ \cite{Pstanford}.
The simplest lattice calculation with an infinitely heavy charmed antiquark and
four light quarks $uuds$,
can easily be done in parallel with the more complicated H calculation
both in the symmetry limit where all light quarks have the same mass
and with $SU(3)$ symmetry breaking.
Comparing the results for these cases may provide
considerable insight into our understanding of the physics of QCD in
multiquark systems even if the pentaquark is not found as
a physical bound state in experiment.
There is therefore interest both in experimental
searches for the pentaquark and in lattice gauge calculations.
However, so far no such lattice calculation has been done or is planned.

    The color-magnetic interaction can also give a molecular-type wave function
extending over a distance large compared with the range of the hyperfine
interaction. Such models have been proposed for the $a_o$ and $f_o$ mesons
\cite{WeinIsg,Lipk83}. A rough estimate of the binding\cite{Lipk83}.
indicates that if the $a_o$ and $f_o$ are really
barely bound $K \bar K$ molecular states\cite{WeinIsg,Lipk83}
the $H$ and Pentaquark should be more strongly bound\cite{Pentaqua} and
be excellent candidates for weakly bound molecular states.

\section{How Good Vertex Detectors Change Signature Requirements}

\subsection{Only background is wrongly-measured events}

Without vertex detectors a pentaquark decay signal appears
against a large combinatorial background. Peaks in a mass spectrum can arise
from statistical fluctuations in the background. Standard statistical
considerations are therefore needed to analyze data.

With good vertex detectors which can distinguish between particles originating
from primary and secondary vertices, every event in which a decay
proton is observed coming from a secondary vertex is evidence for a
particle which decays weakly by proton emission. If its mass is not
consistent with the mass of a known weakly decaying baryon, the event is
either evidence for a new as yet unknown particle or it is a wrong event
incorrectly measured. It cannot be a statistical fluctuation of known physics
if measured correctly. Thus an experiment which can clearly distinguish a proton
coming definitely from a secondary vertex should be considered as an open
search for new weakly decaying baryons and not simply as a pentaquark
search. One might even find new physics beyond the standard model.
This more general framework should be used  both in planning experiments and in
analyzing data from ongoing experiments, rather than considering only the
pentaquark. This point has not been
noted in previous articles on pentaquark searches\cite{PSTGOAR,ZPenta}

Once data are obtained and events remain which pass all cuts
indicating that they have a proton emitted from a secondary vertex,
the main question is not whether this is sufficient evidence for the
pentaquark.
It is rather whether these events are real and indicate evidence for a new
weakly decaying baryon, or whether they are somehow incorrectly measured.
One must investigate all possibilities of incorrect measurement.

\subsection{The Beauty Pentaquark $P_{\bar b suud}$ }

The same considerations apply to baryons containing $b$ quarks, where every
event in which a decay proton is observed coming from a secondary vertex may
also indicate the presence of a new particle. The calculations indicating that
the charmed pentaquark may be bound apply also for analogous states with heavier
antiquarks; e.g. the beauty pentaquark $P_{\bar b suud}$ which are equally
attractive candidates for bound exotics. Many such calculations can be used
directly for the beauty pentaquark by simply replacing the $\bar c$ everywhere
by $\bar b$. The lowest threshold here is $p B_s$ and many of the same analyses
of the $p D_s$ system used for the charmed pentaquark can also be taken over
directly here, including the molecular model in which an off-shell $B_s$ is
bound to a proton. There is, however one essential new ingredient; namely the
existence of two dominant decay modes for the $b$ quark; namely
$b \rightarrow c \bar u d $ and $b \rightarrow c \bar c s $. The second can
give rise to striking signatures in which a proton and a charmonium are
emitted from the same secondary vertex; e.g.
$P_{\bar b suud} \rightarrow p \phi \psi$. Note that in this decay the mass
of the $\phi \psi$ system is below the $B_s$ mass and cannot arise from $B_s$
decay.

\subsection{Tests to distinguish wrongly-measured events}

The mass spectrum offers interesting clues. An anticharmed strange baryon with
a mass above the $D_s - p$ threshold will decay strongly to $D_s - p$
in a primary vertex and not be observed at the secondary vertex. Thus all
events in which a secondary proton is observed but which have a mass above the
$D_s - p$ threshold must be considered as wrong events or really exotic
new physics. The presence of such events which pass all cuts would probably
indicate an unknown systematic error that introduces spurious events without
secondary vertex protons,

A real new weakly-decaying baryon should not only produce a peak in the mass
spectrum at the particle mass, but also a tail below the peak
corresponding to decays in which one or more neutral particles have escaped
detection. Since any weakly-interacting particle also has semileptonic decays
in the standard model, events in the tail containing a muon or electron should
be be expected. These are particularly significant since they confirm the
identification of a weakly decaying particle.

The original suggestion directing the search to particular decay modes like
$p \phi \pi^-$ arose from the necessity to find a signal against a large
statistical combinatorial background. The strategy can be quite different
when the only background is wrongly measured events. One might first
try an extremely stringent cut on the proton to ensure that it really is
a proton that comes from a secondary vertex. This would include events
with all possible numbers of prongs. One can then separate these into
events with even and odd numbers of prongs, which
correspond respectively to decays of neutral and charged particles.
A charged pentaquark with the structure of a $D_s$ bound to a neutron
rather than a proton would be more apt to decay to a final state containing a
neutron rather than a proton, unless the final state baryon is a $\Delta^o$
or $N^{*o}$ decaying to $p \pi^-$. This immediately suggests looking for
$p \pi^-$ resonances in all odd prong events.

The quasi-two-body events now become of interest. The $p\pi^-$ decay
mode, for example would be an impossible signature without vertex
detectors. But once the
proton which definitely comes from a secondary vertex is selected, the unique
energies of the particles provide a striking signal for the new particle. Even
if the efficiency for selecting a secondary proton is only 5\%; i.e.
only one out of every 20 events is really a secondary proton, the
selection of a proton with an energy near the value for a
$p\pi^-$ decay would enhance this signal/noise ratio by a considerable
factor. One would also expect the background from single pions apparently from
the same vertex and having the right energy to be much smaller than the normal
combinatorial background from multiparticle events.

If at this stage the background of presumably wrongly-measured events is
still too large,
the next step is to introduce cuts which hopefully eliminate wrong events
without excessively reducing good events. The effectiveness of a cut can be
estimated by examining the ratio of the events above the $D_s - p$ threshold
to those below the threshold.

If cuts finally eliminate all events above the $D_s - p$ threshold
and others below the threshold remain, the question arises what they might be.
If they are all concentrated in the same mass bin, they might be considered as
evidence for a new particle. If they are scattered over the mass spectrum, they
must be due to some systematic error which has not been eliminated by cuts.
There are also possibilities of a proton from the decay of a known charmed
baryon combined with other particles from a different vertex. There can also be
events from the same new particle with a lower mass because of a decay mode
with an unobserved photon or $\pi^o$, or a semileptonic decay with an
unobserved neutrino and a muon misidentified as a pion.

A real puzzle arises if several events in a single mass bin are observed and
several other events with masses definitely different but too close to the
other mass to be due to the same particle with a missing $\pi^o$. If the events
are not real correctly measured events, the question arises why they only
are found in a very small mass range. If they are all real, they indicate too
many new particles, or some new multiplet not anticipated by theorists.

The ultimate conclusion if a small number of events are found which are not yet
sufficient evidence for a new particle, but cannot be easily dismissed as
due to known systematics, is that more data are needed\cite{Sharon}.

\subsection{Lifetime Considerations}

It is not obvious how the lifetime of the pentaquark may differ from the $D_s$
lifetime. An off-shell $D_s$ bound to a proton would have a longer lifetime
because the transition matrix elements would be essentially the same, but
phase space would be smaller. A more complicated wave function can have factors
working in both directions. There can be color factors inhibiting decay into
a color-singlet baryon and several mesons. There can be new channels opening
that can enhance the decay rate.

However, if the lifetime of the Pentaquark is longer than that of a charmed
meson, another search criterion is opened. One can consider only events which
have a delay greater than several $D_s$ lifetimes. If, for example, the lifetime
of the pentaquark is longer than the $D_s$ lifetime by a factor of 2, the choice
of a minimum delay which reduces the Pentaquark signal by a factor of 10 will
reduce background by a factor of 100.

\section{The Doubly-Strange and Beauty Pentaquarks}

    The doubly strange $P_{\bar c ssud}$ is degenerate with the singly
strange $P_{\bar c suud}$ in the SU(3) flavor symmetry limit and the two states
have equal binding energies relative to the relevant $\Lambda D_s$ and $ND_s$
thresholds. The effects of SU(3) symmetry breaking have been investigated in
detail for the $P_{\bar c suud}$ in the limit where the charmed
quark has infinite mass and its hyperfine interaction energy is neglected. This
limit should be an even better approximation for the beauty quark. We therefore
treat both pentaquarks on the same footing in the remainder of this paper and
use the notation $P_{\bar Q suud}$ and $P_{\bar Q ssud}$ for both, where $Q$
denotes either a $c$ or $b$ quark. We now generalize the treatment of SU(3)
breaking in this limit to apply to both singly and doubly strange pentaquarks
and show that all states have very similar properties and that all may well be
bound.

The stability against breakup of an exotic multiquark system has been
examined by checking whether hyperfine energy can be gained by recoupling the
color and spins of the lowest lying two-hadron threshold
 \cite{Lipisg}.
A variational approach has been used with a wave function in which the two-body
density matrix is the same for all pairs as in a baryon, and the experimental
$N-\Delta$ mass splitting is used to determine the strength of
the hyperfine interaction energy
\cite{Lipisg}.

In our approximation we disregard the heavy antiquark and its small
hyperfine interaction and consider only the classification of
the state of the four quarks coupled to spin zero and a color triplet.
These four quarks in all pentaquark states considered here contain one
quark pair or ``diquark" with the same flavor and one pair with a
different flavor. We use the notation $\ket{d_{21}^S} $ and $\ket{d_{15}^S} $
to denote these quark pair or diquark
states with spin $S$ classified in the symmetric $21$ and the antisymmetric $15$
dimensional representations of the color-spin SU(6).
The Pauli principle requires the quark pair of the same
flavor to be coupled to the antisymmetric $15$.
We now define the classification of the states under consideration
in a conventional notation $\ket{D_6,D_3,S,N}$
\cite{Patera,Sorba}
where $D_6$ and $D_3$ denote the dimensions of the color-spin $SU(6)$ and
color SU(3) representations in which the multiquark states are classified,
$S$ and $N$ denote the total spin and the number of quarks in the system and
primes distinguish between different representations with the same dimension.
$$
\eqalignno{
  \ket{N} &= \ket{70,1,1/2,3} &(WW6a)   \cr
  \ket{\Delta} &= \ket{20,1,3/2,3} &(WW6b)   \cr
  \ket{\Lambda} &= \ket{70,1,1/2,3} &(WW6c)   \cr
  \ket{H} &= \ket{490,1,0,6}        &(WW6d)   \cr
  \ket{d_{21}^0} &= \ket{21,3^*,0,2}  &(WW6e) \cr
  \ket{d_{21}^1} &= \ket{21,6,1,2}  &(WW6f) \cr
  \ket{d_{15}^0} &= \ket{21,6,0,2}  &(WW6g) \cr
  \ket{d_{15}^1} &= \ket{21,3^*,1,2}  &(WW6h) \cr
  \ket{P_{\bar Q suud}}  &= \ket{210,3,0,4}  &(WW6i) \cr
  \ket{P'_{\bar Q suud}} &= \ket{105',3,0,4}  &(WW6j) \cr}
$$
The pentaquark $P_{\bar Q suud}$ is classified in the $210$ of $SU(6)$ to
optimize the hyperfine interaction.
We also define the state $P'_{\bar Q suud}$ in the $105'$ of $SU(6)$.
 It is convenient also to define
the state $\ket{d_{15}^0; d_{21}^0}$ of two spin-zero diquarks with one
diquark classified in the symmetric $21$ and the other in the antisymmetric
$15$ dimensional representation of the color-spin SU(6) and the analogous state
$\ket{d_{15}^1; d_{21}^1}$ of two spin-one diquarks. These two states are
orthogonal linear combinations of the two states $\ket{210,3,0,4}$ and
$\ket{105',3,0,4}$.

    We use the simplified form of the color-spin hyperfine interaction
\cite{Jaffe}
commonly applied to systems containing only quarks and no active antiquarks:
 $$ V = -(v/2)[C_6 - C_3 -(8/3)S(S+1) - 16N]    \eqno(WW7)   $$
    where $v$ is a parameter defining the strength of the interaction,
$C_6$ and $C_3$ denote the eigenvalues of the Casimir operators of
the $SU(6)$ color-spin and $SU(3)$ color groups respectively,
Thia interaction (WW7) is easily evaluated for the states (WW6) by
substituting the eigenvalues of the Casimir operators
\cite{Patera,Sorba}:
$$
\eqalignno{
  C_6(70) &= 66 &(WW8a) \cr
  C_6(20) &= 42 &(WW8b) \cr
  C_6(490)&= 144 &(WW8c) \cr
  C_6(210) &= (304/3) &(WW8d) \cr
  C_6(105') &= (208/3) &(WW8e) \cr
  C_6(21) &= (160/3) &(WW8f) \cr
  C_6(15) &= (112/3) &(WW8g) \cr
  C_3(4) &=(16/3)  &(WW9a)\cr
  C_3(6) &=(40/3)  &(WW9b)\cr
  C_3(8) &= 12. &(WW9c)\cr}
$$
We therefore obtain
$$
\eqalignno{
V(N) = V(\Lambda) = V(d_{21}^0)  &= -8v  &(WW10a)  \cr
V(\Delta) &= 8v  &(WW10b)  \cr
  V(H)      &= -24v &(WW10c)  \cr
  V(d_{15}^0)  &= +4v &(WW10d)  \cr
  V(d_{15}^1)  &= +(8/3)v &(WW10e)  \cr
  V(d_{21}^1)  &= -(4/3)v &(WW10f)  \cr
  V(P_{\bar Q suud}) &= -16v &(WW10g)  \cr
  V(P'_{\bar Q suud}) &= 0 &(WW10h)  \cr
  V(d_{15}^0; d_{21}^0)  &= -4v  &(WW10i)  \cr}
$$
where we have noted that
the hyperfine interaction vanishes between quarks in a spin-zero diquark and
quarks outside the diquark. Thus we can write
$$  V(d_{15}^0; d_{21}^0) =
|\langle{d_{15}^0); d_{21}^0}\ket{210,3,0,4}|^2 V(P_{\bar Q s}) +
|\langle{d_{15}^0); d_{21}^0}\ket{105',3,0,4}|^2 V(P'_{\bar Q s}) = -4v
\eqno(WW11) $$
Substituting eqs. (WW10g) and (WW10h) then gives the result already
noted in ref.\cite{Gignoux}.
$$ |\langle{d_{15}^0; d_{21}^0}\ket{210,3,0,4}|^2 =
|\langle{d_{15}^1; d_{21}^1}\ket{105',3,0,4}|^2 = 1/4
\eqno(WW12a)   $$
$$ |\langle{d_{15}^1; d_{21}^1}\ket{210,3,0,4}|^2 =
|\langle{d_{15}^0; d_{21}^0}\ket{105',3,0,4}|^2 = 3/4
\eqno(WW12b)   $$
The gain in hyperfine interaction for the
$  P_{\bar c suud} $
over the $ND_s$ or $\Lambda D$ threshold (degenerate in this symmetry
limit) was shown to be equal to
the gain for the $H$ over the relevant $\Lambda \Lambda$ threshold
and just half the $\Delta - N$ mass splitting.
$$
\eqalignno{
  M(\Delta)-M(N) &= V(\Delta) - V(N) = 16v  &(WW13a)  \cr
  B(H) &= V(H) - 2V(\Lambda) = -8v =-(1/2) [M(\Delta)-M(N)]
  &(WW13b)  \cr
  B(P_{\bar c suud}) &= V(P_{\bar c suud}) - V(\Lambda) = -8v
  =-(1/2)[M(\Delta)-M(N)]  &(WW13c)  \cr}
$$
where B(X) denotes
the difference in hyperfine energy between the state X and
the relevant threshold. Thus the $  P_{\bar c suud} $ appeared to be
an equally attractive candidate for hyperfine binding as the H
dibaryon \cite{Gignoux}.

     The introduction of
SU(3) symmetry breaking has been shown to reduce the binding of the
$H$ \cite{Rosner}, and
the strange $P_{\bar c suud}$\cite{Gignoux}.
A similar effect occurs for the doubly-strange $P_{\bar Q ssud}$.
This is easily seen by noting that hyperfine binding energy of both the
$  P_{\bar Q suud} $
and the $H$ is reduced by reducing the color-magnetic
interaction of the strange quark. However the strange quark plays no role
in the magnetic interactions of the
$\Lambda \Lambda$,
$\Lambda D$, $N D_s$ and $\Lambda D_s$ final states and their hyperfine
binding energies are unaffected by SU(3) symmetry breaking.

It is convenient to write the broken-$SU(3)$ hyperfine interaction in the form
for a four quark state in which two quarks of the same flavor are coupled to
the $15$ of SU(6) and the remaining two quarks with different flavor are coupled
to the $21$ of SU(6),
$$ V_{br} = a_o V_o + (a_{21}-a_o)  V_{21} + (a_{15}-a_o) V_{15}
 \eqno (WW14) $$
where $V_o$ denotes the values of the hyperfine interaction (WW10i) for the
Pentaquark state $\ket{210,3,0,4}$ in the SU(3) limit for the case where all
the quarks have the nonstrange mass $m_u$. The coefficient $a_o$ is chosen so
that the term $a_o V_o$ contains all the contributions from hyperfine
interactions between the two diquarks with the broken-SU(3) masses. The
remaining two terms give the correction to the hyperfine interaction within
each diquark when the correct masses are used, and the appropriate spin
averages defined by the weighting factors (WW12) are used. For the case where
the wave function $\ket{210,3,0,4}$ is used,
$$ \langle V_{21} \rangle = {{3}\over{4}}\cdot V(d_{21}^1)  +
{{1}\over{4}}\cdot V(d_{21}^0)  =- 3v  \eqno (WW15a) $$

$$ \langle V_{15} \rangle = {{3}\over{4}}\cdot V(d_{15}^1)  +
{{1}\over{4}}\cdot V(d_{15}^0)  =+ 3v  \eqno (WW15b) $$

Thus for the case of the
doubly-strange $P_{\bar c ssud}$,
$$ V_{br}(P_{\bar c ssud}) = {{m_u}\over{m_s}}\cdot V_o
+ {{m_s -  m_u}\over{m_s}}\cdot V_{21}
- {{m_u}\over{m_s}}\cdot {{m_s -  m_u}\over{m_s}}\cdot V_{15}
= (1 - \delta) V_o + \delta V_{21} - (1 - \delta)\delta\cdot  V_{15}
 \eqno (WW16a) $$
where
$$\delta \equiv  {{m_s -  m_u}\over{m_s}}                \eqno (WW16b) $$
is a parameter defined\cite{Rosner} to express the suppression of the strange
quark hyperfine interaction. For our purposes it is sufficient to work to first
order in the SU(3)-breaking perturbation $\delta$. Thus
$$ V_{br}(P_{\bar Q ssud}) \approx
V_o - \delta \cdot (V_o + V_{15} - V_{21}) = -16v + 10 \delta v  \eqno (WW17) $$

The case of the doubly-strange $P_{\bar Q ssud}$ is particularly simple because
the two quarks in each diquark have the same mass even when SU(3) is broken and
the flavor permutation symmetry within each diquark in conserved by the broken
SU(3) interaction.
In the case of the singly-strange $P_{\bar Q suud}$, the two quarks in the
symmetric $21$ diquark have different masses and the broken SU(3) interaction
can have off-diagonal matrix elements which change
the color-spin permutation symmetry of the diquark, and connect the state to a
state in which both diquarks are in the color-spin antisymmetric $15$. This
contribution which introduces a new state outside the two-dimensional Hilbert
space defined by the states $\ket{210,3,0,4}$and $\ket{105',3,0,4}$ has been
neglected in previous calculations\cite{Gignoux}. We follow this example
and neglect these off-diagonal elements which in any case introduce
contributions higher order in $\delta$. In this approximation the hyperfine
interaction between diquarks is given by an average hyperfine interaction..
$$ V_{br}(P_{\bar Q suud}) \approx
\left(1 - {{\delta}\over{2}}\right)\cdot V_o + {{\delta}\over{2}}\cdot V_{15}
- {{\delta}\over{2}} \cdot V_{21} =
V_o - {{\delta}\over{2}} \cdot (V_o - V_{15} + V_{21})= -16v + 11 \delta v
\eqno (WW18) $$
in agreement with the previous result\cite{Gignoux}.
We thus see that the effects of SU(3) symmetry breaking on the doubly strange
pentaquark are roughly equal to those for the singly-strange case.

One must also consider the possibility that the pentaquarks may not be bound but
may be observable as low-lying resonances near threshold in the $D_s p$,
$D_s \Lambda$ $B_s p$ or
$B_s \Lambda$ system. The combinatorial background can be expected to be very
large in the $D_s p$ or $B_s p$ system where many uncorrelated protons can
be present in any
high energy event. In this case the doubly-strange pentaquark may be a more
easily identified candidate than the singly-strange pentaquark. The number of
uncorrelated $\Lambda 's$ can be expected to be very much less than the number
of uncorrelated protons. In either case the pentaquark resonance should appear
as an enhancement near the lower end of phase space.

\section{Decay Modes}

If the pentaquark is a loosely bound $D_s - p$ or $D_s - \Lambda$ molecule,
its decay will resemble that of an off-shell $D_s$. However, it may also have
the structure of a $\bar Q$ and the four-quark state (WW6i),
$$ \ket{P_{\bar Q suud}}  = \ket{3^*,1/2}_{\bar Q} \otimes
\ket{3,0}_{(uuds)}    \eqno (WW19)                              $$
where we have labeled the states by the quantum numbers $\ket{D_3,S}$
and now included the quantum numbers of the $\bar c$.
In the spectator model,
the decay of the pentaquark is described as
$$ \ket{P_{\bar c suud}} \rightarrow  \ket{3^*,1/2}_{(\bar s \bar u d)} \otimes
\ket{3,0}_{(uuds)}    \eqno (WW20)                              $$

There are many ways that this $(3^*,3)$ color
configuration can fragment into a final state of several color-singlet hadrons,
including many states which are not reached in the model of an off-shell
$D_s$ bound to a spectator proton. The lifetime of the $P_{\bar c suud}$ may
therefore be considerably different from that of the $D_s$.
However, even for decay modes like $p \phi \pi^-$ and $p K^{*o} K^-$ which
occur in the $D_s - p$ model, the ratio of the two branching ratios can be
quite different from that for the $D_s$. This can be seen by examining the
initial state (WW19) in which the $\bar c s$ system is not spin zero
as in the $D_s$ but is a linear combination of zero and one, with a factor of
three favoring spin one. It is like a linear combination of $D_s$ and $D^*_s$.

The $D^*_s$ component can decay into vector-pseudoscalar via s-wave in contrast
to the $D_s$ decay which is p-wave. We now show that the s-wave decay has an
additional spin factor of 3 favoring the $\phi \pi$ mode over $ K^{*o} K^-$.

The s-wave decay has no orbital angular momentum in the final four-quark state.
Thus we need consider only spin couplings.
Since the decay is rotationally invariant we can choose the initial state to
be polarized with ``spin up" for convenience.
We now write the color-favored $\bar c \rightarrow \pi^- \bar s $ decay to
include spinology and the combination with the spectator quark
$$ \ket {\bar c_\uparrow} \cdot \ket {s_\uparrow} \rightarrow
\ket{\pi^- \bar s_\uparrow}  \cdot \ket {s_\uparrow} =
\ket{\pi^- }  \cdot \ket {\bar s_\uparrow s_\uparrow} \eqno(WW21a)  $$
Thus, the color-favored $\bar c \rightarrow \pi^- \bar s $ decay  at the
quark level leads to a unique spin coupling of the $\bar  s$ with the spectator
quark and requires the $s \bar s$ state to have the desired spin 1 for the
$\phi-\pi$ decay.

The color suppressed $\bar c  \rightarrow K^{*o} \bar u $ decay written to
include spinology and the combination with the spectator quark is
$$ \ket {\bar c_\uparrow} \cdot \ket {s_\uparrow} \rightarrow
\sqrt{{2\over 3}}\cdot
\ket{K^{*o}_\uparrow \bar u_\downarrow}  \cdot \ket {s_\uparrow} -
\sqrt{{1\over 3}}\cdot
\ket{K^{*o}_\rightarrow \bar u_\uparrow }  \cdot \ket { s_\uparrow}
= $$ $$
= \sqrt{{2\over 3}}\cdot
\ket{K^{*o}_\uparrow }  \cdot \ket { \bar u_\downarrow s_\uparrow} -
\sqrt{{1\over 3}}\cdot
\ket{K^{*o}_\rightarrow } \cdot \ket { \bar u_\uparrow s_\uparrow}
= $$ $$
= \sqrt{{1\over 3}}\cdot \ket{K^{*o}_\uparrow}\cdot \ket {(\bar u s)_{S=0}} +
\sqrt{{1\over 3}}\cdot \ket{K^{*o}_\uparrow}\cdot \ket {(\bar u s)_{S=1}} -
\sqrt{{1\over 3}}\cdot
\ket{K^{*o}_\rightarrow } \cdot \ket { \bar u_\uparrow s_\uparrow}
\eqno(WW21b)  $$
Here the color suppressed $\bar c  \rightarrow K^{*o} \bar u $ decay  at the
quark level is seen to lead to spin couplings of the $\bar  u$ with the
spectator quark which give a probability of (1/3) for the $u \bar s$ state to
have the desired spin 0 for the $ K^{*o} K^-$ decay.

The $D^*_s$ component can also decay purely leptonically into $\mu \nu$ and
$e \nu$ without the helicity suppression factor existing for the pseudoscalar
$D_s$ decay. Since branching ratio for the decay. $D_s \rightarrow \mu \nu$ is
about 1\%, one might expect significantly higher branching ratios for
 $ \ket{P_{\bar c suud}} \rightarrow p \mu \nu$ and
 $ \ket{P_{\bar c suud}} \rightarrow p e \nu$
if the pentaquark has the structure (WW19) of a $\bar c$ and a four-quark
state  (WW6i),

\section{Conclusion and Acknowledgement}

This paper is contributed to a Memorial Volume for Carl Dover, who had always
been interested in searches for exotic hadrons and whom I joined in an
investigation of the physics of H dibaryon searches\cite{PHPROD}. I should like
to acknowledge many stimulating discussions with the members of the Tel Aviv
experimental group\cite{Sharon,ZPenta} and in particular Danny Ashery,
Sharon May-Tal Beck, Gilad Hurvits, Jechiel Lichtenstadt and Murray Moinester.
The results
of the pentaquark search in the Fermilab E791 Collaboration should soon be
available \cite{E791Col} and hopefully indicate the directions for future
searches with better vertex detectors and particle identification.

{\tighten
}

\begin{references}
\bibitem{Moriond} {Harry J. Lipkin,
% New Possibilities for Exotic Hadrons,
in Hadrons, Quarks and Gluons,
Proceedings of the Hadronic Session of the XXIInd Rencontre de Moriond,
Edited by J. Tran Thanh Van, Editions Frontieres, Gif Sur Yvette - France
(1987), p.691}
\bibitem{Houches} {Harry J. Lipkin,
% Frontiers of the Quark Model,
In The Elementary Structure of Matter,
Proceedings of the Workshop, Les Houches, France, 1987
Edited by J.-M. Richard et al, Springer-Verlag (1987) p.24}
\bibitem{LipKEK}{Harry J. Lipkin,
% Hadron Models and Strong Interaction Dynamics
in Hadron '87, Proceedings of the Second International
Conference on Hadron Spectroscopy, KEK Tsukuba, Japan, edited by
Y. Oyanagi, K. Takamatsu and T.Tsuru, KEK Report 87-7 (1987), p.363.}
\bibitem{PANIC}   {Harry J. Lipkin,
% Multiquark Physics - The Great Challenge for Future Directions
% in the Interplays Between Particle and Nuclear Physics,
In Proceedings of
PANIC '87, XI International Conference on Particles and Nuclei,
Nucl. Phys.  A478, 307c (1988)}
\bibitem{Pentaqua}{Harry J. Lipkin,
% New Possibilities for Exotic Hadrons - Anticharmed Strange Baryons
Phys. Lett.  195B, (1987) 484}
\bibitem{Gignoux}{
C. Gignoux, B. Silvestre-Brac and J. M. Richard,
In The Elementary Structure of Matter,
Proceedings of the Workshop, Les Houches, France, 1987
Edited by J.-M. Richard et al, Springer-Verlag (1987) p.42;
Phys. Lett. {\bf B193} (1987) 323}
\bibitem{Pstanford}{Harry J. Lipkin,
% Are There Bound Exotic Anticharmed Strange Baryons ($\bar csuud$ and
% $\bar csudd)$?
In Proceedings of the International Symposium on The Production and Decay
of Heavy Flavors, Stanford (1987) Edited by Elliott D. Bloom and Alfred
Fridman, Annals of the New York Academy of Sciences, Vol. 535 (1988) p.438}
\bibitem{Sivers}{Harry J. Lipkin,
% Hyperfine Interactions, Key to Multiquark Physics,
% Argonne preprint ANL-HEP-CP-88-34,
In Proceedings of the
Topical Conference on Nuclear Chromodynamics, Argonne National Laboratory,
May (1988) Edited by J. Qiu and D. Sivers, World Scientific, Singapore
(1988) p. 260}
\bibitem{PSTGOAR} {Harry J. Lipkin,
%Some Comments on the Possibilities for Searches for the Pentaquark
% ($\bar csuud$),
% Weizmann preprint WIS-90/42/Nov-PH
in Proceedings of the Rheinfels Workshop 1990 on Hadron Mass
Spectrum, St.Goar at the Rhine, Germany, Sept. 3-6, 1990,
Nucl. Phys.  B(Proc. Suppl.) 21 (1991) 258}
\bibitem{Sharon} {S. May-Tal Beck, (for FNAL E791 Collab.)
% Search for the Pentaquark via the $P^0 \rightarrow \phi \pi p$ decay
In Proceedings of the 1994 Annual Meeting of the Division of Particles and
Fields, Albuquerque, N.M., S. Seidel, Ed., World Scientific, (1995) 1177}
\bibitem{E791Col}{E.M. Aitala et al.,
%"Search for the Pentaquark via the $P^0_{{\bar c}s} \rightarrow \phi \pi p$
%Decay" E791 Collaboration,
FERMILAB-Pub-97/118-E and to be published}
\bibitem{Lipk86}{Harry J. Lipkin,
In   Intersections Between Particle and Nuclear Physics,
Proc. Conf. on The Intersections Between Particle and Nuclear
Physics, Lake Louise, Canada, 1986
Edited by Donald F. Geesaman
AIP Conference Proceedings No. 150, p. 657}
\bibitem{LevFran}{E. M. Levin and L. L. Frankfurt, Zh. Eksperim. i.
Theor. Fiz.-Pis'ma Redakt (1965) 105; JETP Letters (1965) 65}
\bibitem{LS}{ H.J. Lipkin and F. Scheck, Phys. Rev. Lett. 16 (1966) 71}
\bibitem{ALS}{G. Alexander, H.J. Lipkin and F. Scheck,
Phys. Rev. Lett. {\bf 17} (1966) 412}
\bibitem{PAQMREV}{Harry J. Lipkin, Physics Letters {B335} (1994) 500}
\bibitem{SakhZel}{
Ya. B. Zeldovich and A.D. Sakharov, Yad. Fiz 4 (1966)395; Sov. J. Nucl. Phys.
4 (1967) 283}
\bibitem{ICHJLmass}{I. Cohen and H. J. Lipkin, Phys. Lett. {93B}, (1980)
56}
\bibitem{HJLMASS}{Harry J. Lipkin, Phys. Lett. {B233} (1989) 446;
Nuc. Phys. A507 (1990) 205c}
\bibitem{DGG}{A. De Rujula, H. Georgi and S.L. Glashow, Phys. Rev. D12
(1975) 147}
\bibitem{Protvino} Harry J. Lipkin, Nucl. Phys.  A478, (1988) 307c
\bibitem{Igal}{Igal Talmi, Phys. Lett.  205B, (1987) 140}
\bibitem{Arima}{A. Arima, K. Yazaki and H. Bohr, \pl \ 183 (1987) 131}
\bibitem{WhyNuc}{H.J. Lipkin, Phys. Lett. 198B (1987) 131}
\bibitem{TriEx}{H.J. Lipkin, Phys. Lett. 45B (1973) 267}
\bibitem{Jaffe}{R. L. Jaffe, Phys. Rev. Lett. 38, (1977) 195}
\bibitem{Rosner}{J. L. Rosner, Phys. Rev. D 33 (1986) 2043}
\bibitem{Thacker}{P. MacKenzie and H. Thacker, Phys. Rev. Letters 65, 2539
(1985)}
\bibitem{Thacker2}{H. Thacker, private communication}
\bibitem{WeinIsg}{John Weinstein and Nathan Isgur,
Phys. Rev. Lett.  {48} (1982) 659;
Phys. Rev.  {D27} (1983) 588}
\bibitem{Lipk83}{Harry J. Lipkin,
Phys. Lett.   {124B} (1983) 509
}
\bibitem{ZPenta}{M. A. Moinester,
D. Ashery, L. G. Landsberg and H. J. Lipkin,
%How to Search for Pentaquarks in High Energy Hadronic Interactions
Zeitschrift fur Physik A 356 (1996) 207}
\bibitem{Lipisg}{
N. Isgur and H. J. Lipkin, Phys. Lett.  99B, 151 (1981).
}
\bibitem{Patera}{ W. G. McKay and J. Patera, Tables of Dimensions, Indices,
and Branching Rules for Representations of Simple Lie Algebras,
(Marcel Dekker, New York, 1981) p. 98}
\bibitem{Sorba}{
H. H\"ogaasen and P. Sorba, Nucl. Phys. B145 (1978) 119
}
\bibitem{PHPROD}{Murray A. Moinester, Carl B. Dover and Harry J. Lipkin,
% On the Possibility of $H$ Dibaryon Production With Energetic Hyperon
% and Meson Beams,
Phys. Rev. C46 (1992) 1082}

\end{references}
\end{document}